\documentclass[12pt]{article}
\usepackage{epsfig}
\usepackage{amssymb}

\usepackage{setspace}

\makeatletter

\begin{document}

 \bigskip \noindent 
 \centerline{\large \bf The classical diffusion-limited Kronig-Penney system}
 
 \bigskip   \bigskip   \bigskip
 
\centerline{\large \bf D. Bar}

\bigskip \bigskip

\begin{abstract}
{\it  We have previously discussed the classical diffusive system of the 
 bounded
one-dimensional multitrap  using the transfer-matrix method which is generally 
applied  for studying the energy spectrum of the 
 unbounded quantum  
Kronig-Penney multibarrier. It was shown, by this method, that for
certain values of the relevant parameters the bounded multitrap array have unity
transmission and a double-peak phase transitional behaviour. We discuss in this
work, using the same transfer matrix method,  the energy  related to the  
diffusion through the unbounded 
 one-dimensional  multitrap  and find that it   may be expressed in two
 entirely different ways with different results and consequences.   
  Also, it is
 shown that, unlike  the barriers in the Kronig-Penney case, 
  the energies at one face of the imperfect trap greatly differ 
   from the energies at the other face of the same trap.  }

\end{abstract}

{\bf \underline{keywords}: Kronig-Penney system, Transfer matrix, Imperfect trap}

{\bf Pacs numbers}: 71.15.Ap, 66.30.-h, 02.10.Yn

\bigskip \bigskip

\protect \section{Introduction}

The remarkable similarity \cite{Mattis,Roepstorff} between the Schroedinger and the
classical diffusion equations have attracted many authors to discuss 
diffusion limited reactions using quantum methods and terminology (see annotated
bibliography in \cite{Mattis}). For example, the same methods and teminology of
transfer matrices \cite{Merzbacher,Tannoudji,Yu},  which are applied  
 \cite{Merzbacher,Tannoudji} for  discussing quantum multibarrier 
potentials,  have been used \cite{Bar1,Bar2} for discussing the one-dimensional
bounded imperfect multitrap system \cite{Smol,Abramson,Noyes,Havlin} 
through which classical particles diffuse. \par 
The imperfect trap,  which was introduced in \cite{Collins}  and further
discussed by others \cite{Budde,Condat,Taitelbaum}, may serve as a model for many physical
situations. For example, one may find applications of it to rotational diffusion
in chemical reactions \cite{Chuang} or to proteins with active sites deep inside
the protein matrix \cite{Nadler} or to infinite lattice traversed by  a random
walker in the presence of an imperfect trap \cite{Budde}. We note that the
discussion of the {\it bounded} one-dimensional multitrap systems have resulted,
for certain values of its parameters   
\cite{Bar1,Bar2},  in finding somewhat unconventional results. Among these one may
count   a unity
transmission of the density of the diffusing particles through the multitrap
array \cite{Bar1}  or the double-peak phase transition recently found
\cite{Bar2}  in such systems. 
\par
An important aspect of the classical one-dimensional multitrap system which was
not fully discussed thus far is when its length tends to infinity. The analogous
quantum infinite multibarrier, which is the Kronig-Penney system
\cite{Merzbacher,Tannoudji,Kittel},  have  been extensively discussed in the
literature by many authors and it is known by its famous  
 band-gap energy spectrum \cite{Merzbacher,Kittel} which is  widely 
applied  in electronics, semiconductors and solid state physics
\cite{Ashcroft}.    \par
We discuss here the  problem of a very large (infinite) one-dimensional
multitrap system  using  the same transfer matrix method  which were applied for
studying the Kronig-Penney multibarrier potential 
\cite{Merzbacher,Tannoudji,Kittel}. We, especially, 
discuss the energy  of the diffusing  particles and apply similar methods 
as those used for studying the enegy spectrum of 
 the quantum Kronig-Penney multibarrier \cite{Kittel,Ashcroft}.  \par 
 By using the transfer matrix method for the unbounded classical
 multitrap system we obtain a  quadratic characteristic
 equation the two solutions of which give rise to two possible expressions 
 for the energy  of  the
 diffusing particles. Each of these two energies has a part which is associated
 with the left hand face of the trap and another, differently expressed, part
 related to the right hand face of it. The different expressions of each of the
 two energies at the left and right hand sides  of the trap causes these
 energies to greatly differ in value at these faces. That is, we show that by
 merely diffusing through the trap the particles's energy enormously changes. 
  All the analytical results are graphically corroborated.
  \par 
    We  note that
the energies of the bounded one-dimensional multitrap system were found in
\cite{Bar2} to have  phase-transitional characteristics for the case in which an
external field was appended to the system. \par
In Section II we  apply the transfer matrix method for introducing and
discussing  the unbounded
one-dimensional multitrap system  as done  in \cite{Bar1,Bar2}. Note that
by using the transfer matrix  formalism we also use its terminology which
usually refers to an $N$-array system \cite{Tannoudji} rather than to an 
infinite  array one. We
remove this finiteness by letting the number of traps $N$ and the 
total length $L$ of the multitrap to become very large. 
In the numerical part we
assign to $N$ and $L$ the  values of 15000  and 20000 (note that in
 \cite{Tannoudji} a finite  multibarrier potential composed of a few hundred
barriers was used as a model for the infinite
Kronig-Penney system).    
In Section
III we discuss  the energy associated with the diffusing particles 
and find, using Appendices A and B,  the
appropriate expressions for it. In Section IV we calculate the energy for 
 some
specific values of its parameters. In Section V we use the analytical results of
Sections III-IV and those of the Appendices A-B for graphically showing the
energies as functions of its variables. We show that these variables have
certain values at which  the corresponding  energy  becomes  
 disallowed  such as, for example,  when it  tends  
 to  become   negative or to assume  very much large positive values. Some 
 analytical expressions and
 derivations are shown in Appendices A-B. We conclude with a brief summary.

\protect \section{Application of the  transfer matrix method for the unbounded 
one-dimensional multitrap system}

 The  one-dimensional imperfect multitrap system is assumed 
to be arranged along
the whole positive $x$ axis and the diffusing particles which pass through it are
supposed to come from the negative side of it. We denote, as in 
\cite{Bar1,Bar2}, 
the total width of all the traps and  the total interval among them by $a$ 
and $b$ respectively where $a$
and $b$  tend to become very much large.  
The ratio of $b$ to $a$ and the total length $a+b$ of the system are denoted
respectively by $c$ and $L$. 
As in \cite{Bar1,Bar2} we may express $a$ and $b$ by $c$ and $L$ as 
$a=\frac{L}{(1+c)}, \ \ \ \ b=\frac{Lc}{(1+c)}$.  
The period of the
multibarrier system which is 
$\frac{L}{N}$   is denoted by $p$.  
We assume that the multitrap system begins at the point
$x=\frac{b}{N}=\frac{pc}{(1+c)}$. \par 
 The initial
and boundary value problem \cite{Rene} which is appropriate for describing the diffusion
through the $N$ imperfect barriers is   \cite{Bar1,Bar2} 
\begin{eqnarray} && 1) \ \rho_t(x,t)=D\rho_{xx}(x,t), \; \; \; t>0,\; \; \; 0<x\le (a+b) \nonumber \\ 
&& 2) \ \rho(x,0)=\rho_0+f(x), \; \; \; 0<x\le (a+b)  \label{e1}  \\ 
&& 3) \ \rho(x_i,t)=\frac{1}{k}\frac{d\rho(x,t)}{dx}|_{x=x_i}, \; \; \;t>0, \; \; \; 1
\le i \le 2N,  \nonumber 
\end{eqnarray}
where $\rho(x,t)$, $\rho_t(x,t)$ and  $\rho_{xx}(x,t)$ denote respectively the
density of the diffusion particles, its first partial derivative with respect to
the time $t$ and its second partial derivative with respect to $x$. The
dissusion constant $D$ is supposed to have two different values; $D_i$ inside
the traps and $D_e$ outside them where  $D_e>D_i$ \cite{Bar1,Bar2}. The value of
$0.5 \frac{cm^2}{sec}$ is the order of magnitude of the diffusion constant at
room temperature and atmospheric pressure (P. 337 in \cite{Reif}). In the
numerical part here we have assigned to $D_e$ and $D_i$ the respective values of
$0.8 \frac{cm^2}{sec}$  and $0.4 \frac{cm^2}{sec}$. The second
equation of the set (\ref{e1}) is the initial condition which is assumed
\cite{Bar1,Bar2} to depend on $x$ through $f(x)$ and on the constant term $\rho_0$.
The third equation of the set (\ref{e1}) is the boundary value condition at the
location of the traps where each trap has a finite width.  
That is, any trap is characterized by the two  points along the $x$
axis where its left and right hand faces are located. The constant $k$ is the
trapping rate (or the imperfection constant) which characterizes the degree of
imperfection  of the  traps where the ideal trap condition is obtained
when $k \to \infty$. The set (\ref{e1}) may be  decomposed into two separate
problems  as follows \cite{Bar1,Bar2} 

\begin{eqnarray} && 1)  \ \rho_t(x,t)=D\rho_{xx}, \; \; \; t>0, \;\; \; 0<x\le (a+b) \nonumber \\ 
&& 2) \ \rho(x,0)=\rho_0,  \; \ \  \ 0<x\le (a+b)  \label{e2}  \\ 
&& 3) \ \rho(x_i,t)=\frac{1}{k}\frac{d\rho(x,t)}{dx}|_{x=x_i}, \; \; \;t>0, \; \;
\; 1 \le i \le 2N  \nonumber 
\end{eqnarray}

\begin{eqnarray} && 1) \  \rho_t(x,t)=D\rho_{xx}(x,t), \; \; \; t>0,\; \; \;0<x\le (a+b) \nonumber \\ 
&& 2) \ \rho(x,0)=f(x),\; \  \; 0<x\le (a+b)  \label{e3}  \\ 
&& 3) \ \rho(x_i,t)=0,\; \; \; t>0, \; \; \;1 \le i \le 2N \nonumber  
\end{eqnarray}

The sets (\ref{e2}) and (\ref{e3}) respectively represent the diffusion 
 through 
$N$ imperfect and $N$ ideal traps  as may be realized from the third
equations of these sets. Following \cite{Rene} one may write the general
solution of the set (\ref{e1}) as \cite{Bar1,Bar2} 
\begin{equation} \rho(x,t)=A\rho_1(x,t)+B\rho_2(x,t),  \label{e4} \end{equation}
where $\rho_1(x,t)$  and $\rho_2(x,t)$ are respectively the solutions of the
problems (\ref{e2}) and (\ref{e3}). Using the method of separating variables
\cite{Rene} one may find the ideal trap solution \cite{Bar1,Bar2} as 
\begin{equation} \label{e5}  \rho_2(x,t)=\sin(\frac{\pi x}{x_i})
e^{-(\frac{tD\pi^2}{x_i^2})}, \ \ \ \ 1 \le i \le 2N \end{equation} 
The solution $\rho_1(x,t)$ of the imperfect trap problem is given by
\cite{Havlin} 
 \begin{eqnarray} &&   \rho_1(x,t)=\rho_0\biggl(erf(\frac{(x-{\grave x_i})}{2\sqrt{Dt}})+
\exp(k^2Dt+k(x-{\grave x_i}))erfc(k\sqrt{Dt}+
\frac{(x-{\grave x_i})}{2\sqrt{Dt}})\biggr), \nonumber \\ 
&&   1 \le i \le 2N    \label{e6} \end{eqnarray} 
where the $erf(x)$ and $erfc(x)$ are respectively the error and complementary error
functions given by 
$erf(x)=\int_0^xe^{-u^2}du$ and $erfc(x)=1-erf(x)=\int_x^{\infty}e^{-u^2}du$.
The ${\grave x_i}$ denote the $2N$ faces of the $N$ traps.  
Using the transfer matrix method, as done in \cite{Merzbacher,Tannoudji} with
respect to the Kronig-Penney potential and in \cite{Bar1,Bar2} with regard to
the bounded multitrap system, one may write the following equation which relates
the two faces of the $j$-th trap 
\begin{equation} \label{e7} 
\left(
\begin{array}{c} A_{2j+1} \\ B_{2j+1} \end{array} \right)=\left[
\begin{array}{cc} T_{11}({\grave x_j}^{left},{\grave x_j}^{right}) 
& T_{12}({\grave x_j}^{left},{\grave x_j}^{right}) \\ T_{21}({\grave
x}_j^{left},{\grave x_j}^{right})
&T_{22}({\grave x_j}^{left},{\grave x_j}^{right}) \end{array} \right]\left(
\begin{array}{c} A_{2(j-1)+1} \\ B_{2(j-1)+1} \end{array} \right),   
\ \ \ \ 1 \le j \le N
\end{equation}
 $A_{2j+1}$ and $B_{2j+1}$ are respectively the imperfect and ideal trap coefficients
 respectively of the $j$-th trap and  $A_{2(j-1)+1}$ and $B_{2(j-1)+1}$ are
 those of the $(j-1)$ trap. The two-dimensional matrix $T^{(j)}$ at the right hand 
 side of Eq (\ref{e7})  relates
the left hand face of the $j$-th trap at ${\grave x_j}^{left}$ to its right 
hand face at 
${\grave x_j}^{right}$ where
${\grave x_j}^{right}>{\grave x_j}^{left}$. The  matrix elements $T_{11}$, $T_{12}$, $T_{21}$ and $T_{22}$ are derived in
details in \cite{Bar1,Bar2} and are given in Appendix A. \par
  For a one-dimensional
 $N$ trap
system, which begins at the point $x=\frac{b}{N}=\frac{pc}{(1+c)}$ and has a period $p$  
   one obtains the general transfer matrix equation  \cite{Bar1,Bar2} 
\begin{eqnarray} 
&& \left(
\begin{array}{c} A_{2N+1} \\ B_{2N+1} \end{array}
\right)=T^{(N)}(p(N-\frac{1}{(1+c)}),pN)T^{(N-1)}(p(N-\frac{(2+c)}{(1+c)}),
p(N-1)), \ldots \nonumber \\ 
&& \ldots T^{(2)}(p(1+\frac{c}{(1+c)}),2p)T^{(1)}(\frac{pc}{(1+c)},p)\left(
\begin{array}{c} A_1 \\ B_1 \end{array}
\right) \label{e8} \end{eqnarray}
Each two-dimensional matrix at the right hand side of the last equation is
denoted in its parentheses by the locations of the left and right hand faces of
its corresponding trap. Thus,  one
may realize, for example, that for an array which begins, as remarked, at the
point $x=\frac{b}{N}=\frac{pc}{(1+c)}$  
the locations of the left and hand side faces of the $N$-th trap
are $x_N^{left}=p(N-\frac{1}{(1+c)})$ and $x_N^{right}=pN$ and those 
 of the first trap are 
 $x_1^{left}=\frac{b}{N}=\frac{pc}{(1+c)}$ and  
$x_1^{right}=\frac{a+b}{N}=p$. Note that,  as
remarked in \cite{Bar1,Bar2},  all the two-dimensinal matrices at the right hand
side of Eq (\ref{e8}) have the same values for $D$, $t$, $L$ and $c$ and differ
by only the values of $x$ along the positive spatial axis. Performing the $N$
products at the right hand side of Eq (\ref{e8}) one may obtains an overall
two-dimensional matrix, denoted ${\cal T}_N$,  whose  elements 
${\cal T}_{N_{11}}$, ${\cal T}_{N_{12}}$, 
${\cal T}_{N_{21}}$ and 
${\cal T}_{N_{22}}$  may be
recursively expressed by  
\begin{eqnarray} &&  {\cal T}_{N_{11}}= {\cal T}_{(N-1)_{11}} 
T_{11}(p(N-\frac{1}{(1+c)}),Np) =\ldots = \prod_{j=1}^{j=N}
T_{11}(p(j-\frac{1}{(1+c)}),jp) \nonumber \\ &&
{\cal T}_{N_{12}}={\cal T}_{(N-1)_{12}}=
\ldots= {\cal T}_{2_{12}}={\cal T}_{1_{12}}=T_{12}=0 
\label{e9} \\ 
&& {\cal  T}_{N_{21}}= {\cal T}_{(N-1)_{21}} T_{22}(p(N-\frac{1}{(1+c)}),Np)+
 {\cal T}_{(N-1)_{11}} T_{21}(p(N-\frac{1}{(1+c)}),Np) 
 \nonumber \\ 
&&  {\cal T}_{N_{22}}= 
{\cal T}_{(N-1)_{22}} T_{22}(p(N-\frac{1}{(1+c)}),Np) =
\ldots=\prod_{j=1}^{j=N}
T_{22}(p(j-\frac{1}{(1+c)}),jp) 
  \nonumber \end{eqnarray} 
Note that whereas ${\cal T}_{N_{11}}$ and 
${\cal T}_{N_{22}}$ are each a one-term expression which is
constructed from $N$ products the element ${\cal  T}_{N_{21}}$ is an 
$N$-term expression each of them is composed of $N$ products. 
  Now, using Eq $(A_2)$ in Appendix A (see also the second of Eqs (\ref{e9}))
  one may calculate 
  the trace $Tr$ and the determinant $Det$ of the two-dimensional
matrix $ T^{(j)}$ at the right hand side of Eq (\ref{e7}) 
\begin{eqnarray}  && Tr(T^{(j)})=T_{11}({\grave x_j}^{left},{\grave x_j}^{right})+
T_{22}({\grave x_j}^{left},{\grave x_j}^{right})
=
T_{11}(p(j-\frac{1}{(1+c)}),pj) + \nonumber \\ && 
+ 
T_{22}(p(j-\frac{1}{(1+c)}),pj) \label{e10} \\
 && Det(T^{(j)})=
T_{11}({\grave x_j}^{left},{\grave x_j}^{right})
\cdot T_{22}({\grave x_j}^{left},{\grave x_j}^{right})=
T_{11}(p(j-\frac{1}{(1+c)}),pj)\cdot \nonumber \\ 
&& \cdot 
T_{22}(p(j-\frac{1}{(1+c)}),pj)   \nonumber \end{eqnarray}
  Using Eqs
(\ref{e10}),  and following the analogous Kronig-Penney case
\cite{Merzbacher,Tannoudji,Kittel},  one  may write  the following quadratic characteristic 
equation  of $T^{(j)}$
\begin{eqnarray} &&y^2-y\cdot Tr(T^{(j)})+Det(T^{(j)})=
y^2-y \cdot (T_{11}(p(j-\frac{1}{(1+c)}),pj) + \nonumber \\ 
&& +
T_{22}(p(j-\frac{1}{(1+c)}),pj))+  
T_{11}(p(j-\frac{1}{(1+c)}),pj) \cdot  \label{e11} \\ && \cdot 
 T_{22}(p(j-\frac{1}{(1+c)}),pj) =0 \nonumber \end{eqnarray}
The two roots  $y_+^{(j)}$ and $y_-^{(j)}$ of the last equation which are the required eigenvalues of 
$T^{(j)}$ are
\begin{equation} \label{e12} y_+^{(j)}=
T_{11}(p(j-\frac{1}{(1+c)}),pj), \ \ \ y_-^{(j)}=
T_{22}(p(j-\frac{1}{(1+c)}),pj) \end{equation}
Now, if the two roots $y_+^{(j)}, \ \ y_-^{(j)}, \ \ 1 \le j \le N$  
are different as for the case here (see Eqs
$(A_1)$ and $(A_4)$ in Appendix
A), the two eigenvectors which correspond to them are linearly independent 
and
we may identify, as for the corresponding quantum Kronig-Penney system
\cite{Merzbacher},  the  initial  values $\left( 
\begin{array}{c} A_1 \\ B_1 \end{array} \right)$  from Eq (\ref{e8})  
with the following two 
eigenvectors \begin{eqnarray} &&  T^{(1)}\left(
\begin{array}{c} A_1^{+} \\ B_1^{+} \end{array}
\right)=y_{+}^{(1)}
 \left(
\begin{array}{c} A_1^{+} \\ B_1^{+} \end{array}
\right)= T_{11}(\frac{pc}{(1+c)},p)  \left(
\begin{array}{c} A_1^{+} \\ B_1^{+} \end{array}
\right) \label{e13}   \\  
&&   T^{(1)}\left(
\begin{array}{c} A_1^{-} \\ B_1^{-} \end{array}
\right)=y_{-}^{(1)}
 \left(
\begin{array}{c} A_1^{-} \\ B_1^{-} \end{array}
\right)= T_{22}(\frac{pc}{(1+c)},p)  \left(
\begin{array}{c} A_1^{-} \\ B_1^{-} \end{array}
\right),   \nonumber       \end{eqnarray}
where  $T^{(1)}$ at the left hand sides of Eqs (\ref{e13})  is the
two-dimensional matrix from  the right hand side of Eq (\ref{e7}) for $j=1$ and
$\frac{pc}{(1+c)}$ and $p$ at the right hand sides of Eqs (\ref{e13}) are, as
mentioned,  the respective locations of the left and  right hand sides of the
first trap.  
For these
$y_{\pm}^{(1)}$ one may identify, as for the Kronig-Penney case
\cite{Merzbacher,Tannoudji},   the coefficients $\left(
\begin{array}{c} A_{2N+1} \\ B_{2N+1} \end{array}
\right)$ in Eq (\ref{e8}) with the two eigenvectors 
\begin{eqnarray} \label{e14} 
&& \left(
\begin{array}{c} A_{2N+1}^+ \\ B_{2N+1}^+ \end{array}
\right)=(y^{(1)}_+)^N\left(
\begin{array}{c} A_1^+ \\ B_1^+ \end{array}
\right) \\ && \left( \begin{array}{c}  A_{2N+1}^- \\ B_{2N+1}^- \end{array}
\right)=(y^{(1)}_-)^N\left(
\begin{array}{c} A_1^- \\ B_1^- \end{array}
\right) \nonumber \end{eqnarray} 
From Eqs $(A_1)$,  $(A_4)$, $(A_5)$ and $(A_7)$  in Appendix A
and from realizing  that the variables ${\grave x_i}$ assume either the value of
${\grave x_j}^{left}$ or ${\grave x_j}^{right}$   (see, for example, the
following discussion before Eq (\ref{e31})) one may see that the quantities 
$y^{(j)}_+, \ \ 1 \le j \le N$  are identical and satisfy 
$y^{(1)}_+ = y^{(2)}_+ =\ldots = y^{(N)}_+$.  The other quantities
   $y^{(j)}_-, \ \ 1 \le j \le N$  can be seen to  slightly differ 
   from each other and one may approximately write $y^{(1)}_- \approx  y^{(2)}_- \approx 
   \ldots \approx  y^{(N)}_-$.   \par
  Considering the limit of an infinite multitrap array which is arranged
along the  whole positive $x$ axis we should demand, as for the Kronig-Penney
case \cite{Merzbacher,Tannoudji},  that as the number of
barriers $N$ tend to $\infty$ the right hand sides of Eqs (\ref{e14}) should
not diverge.  That is, we require 
\begin{eqnarray} && |y_+^{(1)}|=\left|T_{11}(\frac{pc}{(1+c)},p)\right|\le 1   \ \  
  \label{e15} \\ 
&&  |y_-^{(1)}|=\left|T_{22}(\frac{pc}{(1+c)},p)\right|
\le 1  \nonumber \end{eqnarray}
Substituting in the last inequalities for $T_{11}$ and $T_{22}$ from Eqs 
$(A_1)$ and $(A_4)$ of
Appendix A one
obtains \begin{equation} \label{e16} \left|\frac{\alpha(D_e,\frac{pc}{(1+c)},t)
\alpha(D_i,p,t)}{\alpha(D_i,\frac{pc}{(1+c)},t)
\alpha(D_e,p,t)}\right|\le 1  \end{equation}

\begin{equation} \label{e17} \left|\frac{\eta(D_e,\frac{pc}{(1+c)},t)
\eta(D_i,p,t)}{\eta(D_i,\frac{pc}{(1+c)},t)
\eta(D_e,p,t)}\right|\le 1,   \end{equation} 
where $\alpha$ and $\eta$ are given respectively by Eqs $(A_5)$ and 
$(A_7)$ in Appendix A.

\protect \section{The  energy of the diffusing particles in the one-dimensional
multitrap system}

In order to be able to reduce  the inequalities
(\ref{e16})-(\ref{e17}) to calculable expressions  we express the 
parameters $\alpha$ and $\eta$, which
were given by Eqs $(A_5)$ and  $(A_7)$ in  Appendix A, in terms of
the energy $E$ of the diffusing particles. We use for that matter the relevant
expressions  of the energy which were fully derived and discussed  in
\cite{Bar2} for the
 multitrap system.   Thus,  using Eqs (\ref{e4})-(\ref{e6}), we can write 
  the  energy $E$  as  
\begin{eqnarray} && E(D,x,{\grave x_i},t)=\frac{1}{2}\rho v^2=
(\rho(D,x,{\grave x_i},t))\frac{D}{t}=
\biggl(A(x,D)\alpha(D,x,{\grave x_i},t)+
B(x,D)\rho_2(D,x,{\grave x_i},t)\biggr)\cdot \nonumber \\ && \cdot \frac{D}{t} =   
\biggl\{A(x,D)\biggl(erf(\frac{(x-{\grave x_i})}{2\sqrt{Dt}})+  
\exp(k^2Dt+k(x-{\grave x_i}))  erfc(k\sqrt{Dt}+  
\frac{(x-{\grave x_i})}{2\sqrt{Dt}})\biggr)+ \nonumber \\ && + 
 B(x,D)\sin(\frac{\pi x}{{\grave x_i}})
\exp(-\frac{Dt\pi^2}{{\grave x_i}^2})\biggr\}\cdot \frac{D}{t},   
  \ \ \  
 i=1, 2, \ldots 2N, \ \ \ \ t>0,   \label{e18}  
\end{eqnarray}
  where $v$ is the average diffusion  velocity  $v=\sqrt{\frac{2D}{t}}$ which is
  derived from the classical one-dimensional diffusion equation 
  for any finite
  $t$ (see, for example P. 91 in \cite{Openu}).  The  variables  ${\grave x_i}$ denote  
  the locations on the $x$ axis of  the
$2N$ faces of the $N$ traps (see  the
solutions (\ref{e5})-(\ref{e6})  of the respective  ideal and 
imperfect trap problems (\ref{e3}) and (\ref{e2})).  
 The  imperfect  
and ideal trap coefficients $A(x,D)$ and $B(x,D)$ are   
 numerically found for the $2N$  faces of the $N$ traps 
 $x={\grave x_j}, \ \ \ 
 j=1, 2, \ldots 2N$ \cite{Bar1,Bar2}. 
   That is,  for each $j$-th trap,  one may find, using the transfer matrix
   method, the four pairs (1) $A({\grave x_j}^{left},D_i), B({\grave
   x_j}^{left},D_i)$, (2) $A({\grave x_j}^{right},D_i), B({\grave
   x_j}^{right},D_i)$,  (3) $A({\grave x_j}^{left},D_e), B({\grave
   x_j}^{left},D_e)$, (4) $A({\grave x_j}^{right},D_e), B({\grave
   x_j}^{right},D_e)$. The first  pair denotes 
    the ideal and imperfect trap
   coefficients {\it inside} the $j$-th trap at its left  hand face. 
   The second  pair denotes  
    these 
   coefficients {\it inside} the $j$-th trap at its right  hand face.   
   The third and fourth pairs  denote these coefficients {\it outside} 
   the $j$-th trap 
   at its left and right hand faces.  Note that these coefficients, as well as
   the variables  ${\grave x_j}^{left}$ and ${\grave x_j}^{right}$, are not
   independent of  each other. First, one may realize  (see the discussion after
   Eq (\ref{e8}))  that  ${\grave x_j}^{left}$ and ${\grave x_j}^{right}$ are
   given by   ${\grave x_j}^{left}=p(j-\frac{1}{(1+c)}), 
\ \ {\grave x_j}^{right}=pj, \ \ 1 \le j \le N$ so that they are 
related by 
 \begin{equation} \label{e19}  {\grave x_j}^{left}={\grave x_j}^{right}-\frac{p}{(1+c)}, 
  \ \ \ 1 \le j \le N \end{equation}
  Second, the transfer matrix method relates the
  former coefficients of the $j$-th trap among themselves and also with 
  those of the $(j+1)$-st trap as  \cite{Bar1} 
  \begin{eqnarray} && A({\grave x_j}^{left},D_i)=A({\grave x_j}^{right},D_i),   
  \ \ \ A({\grave x_j}^{right},D_e)=A({\grave x_{(j+1)}}^{left},D_e) \label{e20} 
  \\ && B({\grave x_j}^{left},D_i)=B({\grave x_j}^{right},D_i), 
  \ \ \ B({\grave x_j}^{right},D_e)=B({\grave x_{(j+1)}}^{left},D_e) \nonumber 
  \end{eqnarray} 
  As one may realize the ${\grave x_i}$ from Eq (\ref{e18}) does not have to
  coincide with ${\grave x_j}$. That is, although for the same $j$-th trap 
  each of  
  ${\grave x_j}$ and ${\grave x_i}$ denote its two faces,   ${\grave x_j}$
  may, for a specific context, refers to its left hand face in which 
  case it is written as 
  ${\grave x_j}^{left}$ whereas  ${\grave x_i}$ may refers in this context  
  to its right hand face and is written as 
  ${\grave x_j}^{right}$. \par 
Now,  analogously to the Kronig-Penney 
case \cite{Merzbacher,Tannoudji,Kittel},  one may turn   the inequalities  at the right hand 
sides  of (\ref{e16})-(\ref{e17}) to equalities.  In such case the left hand
sides of  (\ref{e16})-(\ref{e17}) are equated to $\cos(\kappa p)$ 
 where $\kappa$  is a  real parameter and $p$ is the 
 period of the multitrap which
 is $p=\frac{L}{N}$.    In accordance with 
 the analogous procedure of the
 Kronig-Penney case \cite{Merzbacher,Tannoudji,Kittel} the two eigenvalues 
 of the
 characteristic equation are related to the same parameter.   Note that 
   even if
 one relates the two eigenvalues to different parameters he will obtain 
 the same
 following expressions (\ref{e31})-(\ref{e34}) for the energies  each of which
 depends on only  one parameter.  \par
   We  use in the following  the transfer
matrix principal  property in which the density (and its derivative) at 
the two
sides  of any face of each trap are equal \cite{Merzbacher,Tannoudji,Bar1}.  
This 
may be expressed, for example, for the left hand face of the $j$-th trap as   
\begin{eqnarray} &&  
\rho(D_e,{\grave x_j}^{left},{\grave x_i},t)=A({\grave x_j}^{left},De)\alpha(D_e,{\grave 
x_j}^{left},{\grave x_i},t)+B({\grave x_j}^{left},D_e)\rho_2(D_e,{\grave
x_j}^{left},{\grave x_i},t)
= \nonumber \\ && =\rho(D_i,{\grave x_j}^{left},{\grave x_i},t)= 
 A({\grave x_j}^{left},D_i)\alpha(D_i,{\grave x_j}^{left},{\grave x_i},t)+B({\grave
 x_j}^{left},D_i)\rho_2(D_i,{\grave x_j}^{left},{\grave x_i},t) \label{e21}  
 \end{eqnarray}
  Substituting from Eq (\ref{e18})  for the 
 $\alpha$'s in (\ref{e16}) and from Eqs (\ref{e5}), (\ref{e18}) and 
 $(A_7)$   
   for the $\eta$'s in (\ref{e17}) 
 one obtains after  
 equating the left hand sides of (\ref{e16})-(\ref{e17})  to $\cos(\kappa p)$ 
   (see the discussion after Eq
 (\ref{e20}))   
  \begin{eqnarray} &&  \frac{ \left( \frac{E(D_e,\frac{pc}{(1+c)},
  {\grave x_i},t)t}{A(\frac{pc}{(1+c)},D_e)D_e}-
  \frac{B(\frac{pc}{(1+c)},D_e)\rho_2(D_e,\frac{pc}{(1+c)},
  {\grave x_i},t)}{A(\frac{pc}{(1+c)},D_e)}\right) \cdot \left( 
  \frac{E(D_i,p,
  {\grave x_i},t)t}{A(p,D_i)D_i}-
  \frac{B(p,D_i)\rho_2(D_i,p,
  {\grave x_i},t)}{A(p,D_i)}\right)}{\left( 
  \frac{E(D_i,\frac{pc}{(1+c)},
  {\grave x_i},t)t}{A(\frac{pc}{(1+c)},D_i)D_i}-
  \frac{B(\frac{pc}{(1+c)},D_i)\rho_2(D_i,\frac{pc}{(1+c)},
  {\grave x_i},t)}{A(\frac{pc}{(1+c)},D_i)}\right) \cdot \left( 
  \frac{E(D_e,p,
  {\grave x_i},t)t}{A(p,D_e)D_e}-
  \frac{B(p,D_e)\rho_2(D_e,p,
  {\grave x_i},t)}{A(p,D_e)}\right)} = \nonumber 
  \\ && = \cos(\kappa p) \label{e22}
  \end{eqnarray}

\begin{eqnarray} &&  \frac{ \left( \frac{E(D_e,\frac{pc}{(1+c)},
  {\grave x_i},t)t}{B(\frac{pc}{(1+c)},D_e)D_e}-
  \frac{A(\frac{pc}{(1+c)},D_e)\alpha(D_e,\frac{pc}{(1+c)},
  {\grave x_i},t)}{B(\frac{pc}{(1+c)},D_e)}\right) 
  \cdot \left( 
  \frac{E(D_i,p,
  {\grave x_i},t)t}{B(p,D_i)D_i}-
  \frac{A(p,D_i)\alpha(D_i,p,
  {\grave x_i},t)}{B(p,D_i)}\right)}
  { \left( 
  \frac{E(D_i,\frac{pc}{(1+c)},
  {\grave x_i},t)t}{B(\frac{pc}{(1+c)},D_i)D_i}-
  \frac{A(\frac{pc}{(1+c)},D_i)\alpha(D_i,\frac{pc}{(1+c)},
  {\grave x_i},t)}{B(\frac{pc}{(1+c)},D_i)}\right) 
  \cdot \left( 
  \frac{E(D_e,p,
  {\grave x_i},t)t}{B(p,D_e)D_e}-
  \frac{A(p,D_e)\alpha(D_e,p,
  {\grave x_i},t)}{B(p,D_e)}\right) } 
  = \nonumber 
  \\ && = \cos(\kappa p) \label{e23}
  \end{eqnarray}
  The functions $\rho_2$ and  $\alpha$  are given respectively 
  by Eqs (\ref{e5}) and $(A_5)$ in Appendix A and  use is made of the relation
  $\rho_2(D,x,{\grave x_i},t)=-\frac{{\grave x_i}}{\pi}\eta(D,{\grave
  x_i},t)\sin(\frac{\pi x}{{\grave x_i}})$  obtained by comparing  Eq  
  (\ref{e5}) with Eq $(A_7)$ in Appendix A. 
   The sine function and the factor   $\frac{{\grave x_i}}{\pi}$ 
   which do not depend on the diffusion constants $D_i$ and $D_e$ 
     are cancelled in Eq (\ref{e23}).
  As realized from the last equations there are four energies related to the 
  trap;  $E(D_e,\frac{pc}{(1+c)},
  {\grave x_i},t)$, $E(D_e,p,
  {\grave x_i},t)$, $E(D_i,\frac{pc}{(1+c)},
  {\grave x_i},t)$, and $E(D_i,p,
  {\grave x_i},t)$.  
 But,  as seen,     one may reduce the
 number of the energies related to each trap to two since 
 using Eqs (\ref{e18}) and  (\ref{e21}) one may obtain the following expressions
 which relate the energies 
 at the two
 sides of each trap  \begin{eqnarray}  
 && E(D_e,{\grave x_j}^{left},{\grave x_i},t)\frac{t}{D_e}=
 E(D_i,{\grave x_j}^{left},{\grave x_i},t)\frac{t}{D_i}, \ \ 1 \le j \le N  \label{e24} \\ 
&&  E(D_e,{\grave x_j}^{right},{\grave x_i},t)\frac{t}{D_e}=
 E(D_i,{\grave x_j}^{right},{\grave x_i},t)\frac{t}{D_i}, \ \ 1 \le j \le  N \nonumber 
 \end{eqnarray}
  In the following  we  use Eqs (\ref{e22})-(\ref{e24}) for finding the 
  two energies  $E(D_e,p,
  {\grave x_i},t)$ and   $E(D_e,\frac{pc}{(1+c)},
  {\grave x_i},t)$ which are respectively the energies at the right and left
  hand faces of the trap.  
  Thus,  using  Eqs (\ref{e24}) we may rewrite  Eqs (\ref{e22})-(\ref{e23}) 
   as follows 
\begin{eqnarray}  &&  \frac{ \scriptstyle \left( E(D_e,\frac{pc}{(1+c)},
  {\grave x_i},t)t-D_eB(\frac{pc}{(1+c)},D_e)\rho_2(D_e,\frac{pc}{(1+c)},
  {\grave x_i},t)\right) \cdot \left( 
  E(D_e,p,
  {\grave x_i},t)t-D_eB(p,D_i)\rho_2(D_i,p,
  {\grave x_i},t)\right)}{ \scriptstyle \left( 
  E(D_e,\frac{pc}{(1+c)},
  {\grave x_i},t)t-D_eB(\frac{pc}{(1+c)},D_i)\rho_2(D_i,\frac{pc}{(1+c)},
  {\grave x_i},t)\right) \cdot \left( 
  E(D_e,p,
  {\grave x_i},t)t-D_eB(p,D_e)\rho_2(D_e,p,
  {\grave x_i},t)\right)}  =  \nonumber \\ 
   && = \frac{A(\frac{pc}{(1+c)},D_e)}
  {A(p,D_e)}\cos(\kappa p) \label{e25}
  \end{eqnarray} 
  
  \begin{eqnarray}  &&  \frac{ \scriptstyle \left( E(D_e,\frac{pc}{(1+c)},
  {\grave x_i},t)t-D_eA(\frac{pc}{(1+c)},D_e)\alpha(D_e,\frac{pc}{(1+c)},
  {\grave x_i},t)\right) \cdot \left( 
  E(D_e,p,
  {\grave x_i},t)t-D_eA(p,D_i)\alpha(D_i,p,
  {\grave x_i},t)\right)}{ \scriptstyle \left( 
  E(D_e,\frac{pc}{(1+c)},
  {\grave x_i},t)t-D_eA(\frac{pc}{(1+c)},D_i)\alpha(D_i,\frac{pc}{(1+c)},
  {\grave x_i},t)\right) \cdot \left( 
  E(D_e,p,
  {\grave x_i},t)t-D_eA(p,D_e)\alpha(D_e,p,
  {\grave x_i},t)\right)}  =  \nonumber \\ 
   && = \frac{B(\frac{pc}{(1+c)},D_e)}
  {B(p,D_e)}\cos(\kappa p) \label{e26}
  \end{eqnarray} 
  
The two  quadratic equations (\ref{e25})-(\ref{e26}) 
were simultaneously solved in Appendix B for the  energies 
$E(D_e,p,
  {\grave x_i},t)$ and  $E(D_e,\frac{pc}{(1+c)},
  {\grave x_i},t)$ and two separate solutions were found for each (see Eqs
  $(B_6)$-$(B_9)$ in Appendix B). For  $E(D_e,p,
  {\grave x_i},t)$ we find the two solutions of 
 
 \begin{equation} E^{+}(D_e,p,{\grave
x_i},t)=\frac{t(X_1X_4-X_5)Y_3-(X_1X_2X_4-X_3X_5)Y_1}
{(tX_3-tX_1X_2)Y_1-t^2(1-X_1)Y_3} \label{e27} \end{equation} 
\begin{equation} E^{-}(D_e,p,{\grave
x_i},t)=\frac{Y_2}{Y_1}
\label{e28} \end{equation} 
And for  $E(D_e,\frac{pc}{(1+c)},
  {\grave x_i},t)$ we find the two solutions of 
\begin{equation}   E^{+}(D_e,\frac{pc}{(1+c)},{\grave
x_i},t)=\frac{Y_3}{Y_1} \label{e29}   \end{equation}
\begin{eqnarray} &&  E^{-}(D_e,\frac{pc}{(1+c)},{\grave
x_i},t)= \label{e30} \\
&&=\frac{ \scriptstyle (X_3-X_1X_2)((tX_3-tX_1X_2)Y_2+
(X_1X_2X_4-X_3X_5)Y_1)-(1-X_1)(t^2(X_3-X_1X_2)(Y_4-Y_5)+t(X_1X_2X_4-X_3X_5)Y_3)}
{ \scriptstyle
(X_3-X_1X_2)((t^2(1-X_1)Y_2+t(X_1X_4-X_5)Y_1)-(1-X_1)(t^3(1-X_1)(Y_4-Y_5)+
t^2(X_1X_4-X_5)Y_3)},  \nonumber 
\end{eqnarray} 
where the quantities $X_1, \ \ X_2, \ \ X_3, \ \ X_4, \ \ X_5$ and  
$Y_1, \ \ Y_2, \ \ Y_3, \ \ Y_4, \ \ Y_5$   are given respectively by Eqs
$(B_1)$ and $(B_4)$ in Appendix B. 
The  energies $E^{+}(D_e,p,{\grave
x_i},t)$ and $E^{-}(D_e,p,{\grave
x_i},t)$ from  Eqs (\ref{e27})-(\ref{e28}) are for  
 the right hand side of the  trap and the energies 
 $E^{+}(D_e,\frac{pc}{(1+c)},{\grave
x_i},t)$ and $E^{-}(D_e,\frac{pc}{(1+c)},{\grave
x_i},t)$ from Eqs (\ref{e29})-(\ref{e30}) are for the left hand side of it. \par
In the former expressions of  the energies  the variable  
${\grave x_i}$ must  coincides with either $\frac{pc}{(1+c)}$ or 
$p$.   
Thus,  when ${\grave x_i}=\frac{pc}{(1+c)}$  one have to discard the 
solutions  $E^{-}(D_e,\frac{pc}{(1+c)},\frac{pc}{(1+c)},t)$ and  
$E^{-}(D_e,p,\frac{pc}{(1+c)},t)$ since in this case one 
obtains from Eq (\ref{e5}) and from Eqs 
$(B_1)$ in Appendix B $X_2=X_3=0$.   In this case the 
 energy $E^{-}(D_e,\frac{pc}{(1+c)},\frac{pc}{(1+c)},t)$ from  
 Eq (\ref{e30}) vanishes whereas the energy  
$E^{-}(D_e,p,\frac{pc}{(1+c)},t)$ from Eq (\ref{e28}) remains at the value of 
$\frac{Y_2}{Y_1}$.    This implied the unreasonable conclusion that
the passing particles have no energies at the left hand face of the  trap 
before they diffuse through it whereas at the right hand face of it they have
nonvanishing unaccountable energies. Thus, for  
${\grave x_i}=\frac{pc}{(1+c)}$  only the energies 
$E^{+}(D_e,p,\frac{pc}{(1+c)},t)$ and 
$E^{+}(D_e,\frac{pc}{(1+c)},\frac{pc}{(1+c)},t)$  must be considered and 
they are given by 
\begin{equation} \label{e31} E^{+}(D_e,p,
\frac{pc}{(1+c)},t)= \frac{(X_1X_4-X_5)}{t(X_1-1)} \end{equation} 
\begin{equation} \label{e32} E^{+}(D_e,\frac{pc}{(1+c)},
\frac{pc}{(1+c)},t)= \frac{Y_3}{Y_1} \end{equation} 
 The second case is that of 
${\grave x_i}=p$  for which one obtains from Eq (\ref{e5})
and from Eqs 
$(B_1)$ in Appendix B  $X_4=X_5=0$.   In this case   the energy 
$E^{+}(D_e,p,p,t)$ from Eq (\ref{e27}) vanishes whereas the energy 
$E^{+}(D_e,\frac{pc}{(1+c)},p,t)$ from Eq (\ref{e29}) remains in the value of $\frac{Y_3}{Y_1}$. 
This also could not be accepted since it
implied that the particles suddenly stop diffusing after the first trap 
whereas we are
concerned with the diffusion along the entire multitrap system. 
Thus, the energies 
$E^{+}(D_e,p,p,t)$  and 
$E^{+}(D_e,\frac{pc}{(1+c)},p,t)$ 
 must be discarded and we have to take into
account only the energies  
$E^{-}(D_e,p,p,t)$ 
 and $E^{-}(D_e,\frac{pc}{(1+c)},p,t)$  
 which are given by 
\begin{equation} \label{e33} E^{-}(D_e,\frac{pc}{(1+c)},p,t)=\frac{(X_3-X_1X_2)}{t(1-X_1)} 
\end{equation}
\begin{equation} \label{e34} E^{-}(D_e,p,p,t) =\frac{Y_2}{Y_1} \end{equation} 
If $c$ becomes very large so that $c>>1$ one may realize from Eqs $(B_1)$ in
Appendix B   and Eqs  (\ref{e5}), 
 (\ref{e31}) and (\ref{e33}) that the energies  
 $E^{+}(D_e,p,\frac{pc}{(1+c)},t)$ and 
  $E^{-}(D_e,\frac{pc}{(1+c)},p,t)$    tend to zero.

\section{Calculation  of the energies  (\ref{e31})-(\ref{e34})  
for specific values of $\kappa p$.  } 

The expressions (\ref{e31})-(\ref{e34})  for the energies 
$E^{+}(D_e,p,\frac{pc}{(1+c)},t)$, 
$E^{+}(D_e,\frac{pc}{(1+c)},\frac{pc}{(1+c)},t)$,  
$E^{-}(D_e,\frac{pc}{(1+c)},p,t)$, 
$E^{-}(D_e,p,p,t)$ should now be evaluated as functions of $\kappa
 p$ as done for the quantum Kronig-Penney system
 \cite{Merzbacher,Tannoudji,Kittel}. 
 But before doing that we show that the expressions (\ref{e31})-(\ref{e34}) 
  become  simplified for certain values of $\kappa p$. 
 Thus, for $\kappa p=\frac{\pi}{2}+n\pi, \ \ n=0, 1, 2, \ldots$ one have 
 $\cos(\kappa p)=0$ and from  Eqs $(B_4)$ and the first of Eqs  
 $(B_1)$ in Appendix B 
 we have $Y_5=X_1=0$  and also the
 second terms of $Y_1$, $Y_2$ and $Y_3$   vanish as well. 
  Thus,   the energies (\ref{e31})-(\ref{e34}) become 
  \begin{equation} E^{+}_{(\cos(\kappa p)=0)}(D_e,p,
  \frac{pc}{(1+c)},t)=
 \frac{X_5}{t}=
 \frac{D_eB(p,D_i)\rho_2(D_i,p,\frac{pc}{(1+c)},t)}{t} 
  \label{e35} \end{equation} 
 
 \begin{eqnarray} &&
 E^{+}_{(\cos(\kappa p)=0)}(D_e,\frac{pc}{(1+c)},
 \frac{pc}{(1+c)},t)
 =\frac{D_eA(\frac{pc}{(1+c)},D_e)\alpha(D_e,\frac{pc}{(1+c)},\frac{pc}{(1+c)},t)}{t}=
 \frac{D_eA(\frac{pc}{(1+c)},D_e)}{t} \cdot \nonumber \\ &&
\cdot 
\exp(k^2D_et)erfc(k\sqrt{D_et})  \label{e36}  
\end{eqnarray}

\begin{equation}  
E^{-}_{(\cos(\kappa p)=0)}(D_e,\frac{pc}{(1+c)},
p,t)=\frac{X_3}{t}=
 \frac{D_eB(\frac{pc}{(1+c)},D_e)\rho_2(D_e,\frac{pc}{(1+c)},p,t)}{t}  
 \label{e37}  \end{equation} 

\begin{eqnarray} &&
 E^{-}_{(\cos(\kappa p)=0)}(D_e,p,p,t)
 =\frac{D_eA(p,D_i)\alpha(D_i,p,p,t)}{t}=\frac{D_eA(p,D_i)}{t} 
 \cdot \nonumber \\ &&
\cdot 
\exp(k^2D_it)erfc(k\sqrt{D_it}) \label{e38}  
\end{eqnarray}
For obtaining Eqs (\ref{e35}) and (\ref{e37}) we respectively use the 
fifth and
third  of Eqs $(B_1)$ in Appendix B and for Eqs 
(\ref{e36}) and (\ref{e38}) we respectively use the third and second of Eqs
$(B_4)$ in Appendix B. Use is also made of Eq $(A_5)$ in Appendix A and
the first of Eqs $(B_4)$ in Appendix B. 
 \par 
Another  kind of points
which draw special attention 
is  $\kappa p=arc(\cos(\frac{A(p,D_e)}{A(\frac{pc}{(1+c)},D_e)}))$ 
 for which one obtains from the first of Eqs
$(B_1)$ in Appendix B $X_1=1$.  At these points   the energies 
$E^{+}(D_e,p,
 \frac{pc}{(1+c)},t)$ and 
 $E^{-}(D_e,\frac{pc}{(1+c)},
 p,t)$ from Eqs (\ref{e31}) and (\ref{e33}) 
 must be discarded since they tend to infinity and this can not be accepted on
 physical grounds.
   Note that using the transfer matrix method one may conclude that 
    the ideal traps coefficients   always satisfy 
    $B({\grave x_j}^{right},D_e)>B({\grave x_j}^{left},D_e)$ and so the division
   $\frac{B(p,D_e)}
{B(\frac{pc}{(1+c)},D_e)}$ is  greater than unity which implies that  the
quantity $Y_1$,  as defined by the first of Eqs $(B_4)$ in Appendix B,   is
always positive. 
This  determines, as will be shown,  the values and the graphical form 
 of the energies 
$E^{+}(D_e,\frac{pc}{(1+c)},
 \frac{pc}{(1+c)},t)$ and $E^{-}(D_e,p,
 p,t)$ from Eqs (\ref{e32}) and (\ref{e34}).

 \section{ The energies  as functions of $\kappa
p$, $c$, $k$, and $t$. } 

As seen from Eqs $(B_1)$ and $(B_4)$ in Appendix B, $(A_5)$ in
Appendix A and  from Eqs 
(\ref{e5})-(\ref{e6})  the energies
(\ref{e31})-(\ref{e34}) critically depend upon the ratio $c$, the trapping rate
$k$ and the time $t$.  Also, one may conclude from the analytical form of the 
expressions  (\ref{e31})-(\ref{e34}) 
and from the Panels of Figures 1-4 that the energy 
$E^{+}(D_e,p,
 \frac{pc}{(1+c)},t)$ from Eq (\ref{e31}) corresponds to 
  $E^{-}(D_e,\frac{pc}{(1+c)},
 p,t)$ from Eq (\ref{e33}) and 
 $E^{+}(D_e,\frac{pc}{(1+c)},
 \frac{pc}{(1+c)},t)$ from Eq (\ref{e32}) corresponds to 
 $E^{-}(D_e,p,
 p,t)$ from Eq (\ref{e34}). That is, 
 $E^{+}(D_e,p,
 \frac{pc}{(1+c)},t)$ and $E^{-}(D_e,\frac{pc}{(1+c)},
 p,t)$ are expressed only by the ideal trap
 expressions  from Eqs $(B_1)$ in Appendix B and 
 $E^{+}(D_e,\frac{pc}{(1+c)},
 \frac{pc}{(1+c)},t)$ and $E^{-}(D_e,p,
 p,t)$ are given only by the imperfect trap
 expressions from Eqs $(B_4)$ of Appendix B.  The corresponding  energies 
 $E^{+}(D_e,p,
 \frac{pc}{(1+c)},t)$ and $E^{-}(D_e,\frac{pc}{(1+c)},
 p,t)$ as functions of $\kappa p$ are characterized with a behaviour 
 which causes  them 
  to  abruptly change their values in a rather jumpy and discontinuous  way 
  (see, for example,  the Panels of Figures 
  1-2). As one may assume  these abrupt changes in the energies  (\ref{e31}) and
  (\ref{e33}) are  related to the values of $\kappa p$ for which 
   $(X_1)$ in their denominator is close to $ 1$. 
   Note that    the  nonzero values of these energies may be negative in 
   which case they 
   can not
  represent real energies   since we discuss here only kinetic  energies 
   as realized  from Eq (\ref{e18}).   The second
  corresponding pair of energies $E^{+}(D_e,\frac{pc}{(1+c)},
 \frac{pc}{(1+c)},t)$ and $E^{-}(D_e,p,
p,t)$ are characterized as steeply increasing with  $\kappa p$ for very 
small values of it and at  $\kappa p \approx 0.05$ they
 become constant (as functions of $\kappa p$)  for all  $\kappa p > 0.05$  
 (see, for example, 
   the Panels of Figures 3 and 4). Also, in contrast to the former pair, 
   these energies are always positive.  \par
 From the Panels of Figures 1-4 one may realize that, for the same values
 of $c$, $k$ and $t$,  the  nonzero values of the 
 energy  
 $E^{-}(D_e,\frac{pc}{(1+c)},
 p,t)$  are generally greater  by several orders of magnitude 
 from the other three energies (see, for example,  Panels 1-2 of Figure 4 
 which
 show a giant difference of $10^{44}$ between $E^{-}(D_e,\frac{pc}{(1+c)},
 p,t)$ and $E^{+}(D_e,p,\frac{pc}{(1+c)},t)$). 
  Thus, although,  as mentioned,  the energies $E^{-}(D_e,p,
 p,t)$ and $E^{-}(D_e,\frac{pc}{(1+c)},
 p,t)$ 
 correspond in analytical expressions and graphical  form 
 as functions of $\kappa p$ to 
  the respective energies  
 $E^{+}(D_e,\frac{pc}{(1+c)},
 \frac{pc}{(1+c)},t)$ and $E^{+}(D_e,p,
 \frac{pc}{(1+c)},t)$ they greatly differ in 
  value.  This may be seen from Panels 1 and 2 of
 Figure 1 which, respectively, show the energies $E^{-}(D_e,\frac{pc}{(1+c)},
 p,t)$ and $E^{+}(D_e,p,
 \frac{pc}{(1+c)},t)$ as functions of $\kappa p$ using  the same 
 20 different values of the ratio $c$  for each Panel and  the same
 $t$ and  $k$ for all the graphs shown. The 20 values of
 $c$ are $c=0.1 +n\cdot 0.4, \ \ \ n=0, 1, 2, \ldots 19$ and the values of 
  $t$ and
  $k$ for all the graphs shown in the two Panels are 
 $t=k=2$. Note that although  the energies shown in Panels 1 and 2
 look similar as functions of $\kappa p$ the nonzero values of these 
 energies differ by as much as $10^4$.   
 As mentioned, we should consider only the zero or the 
 positive parts of the graphs as
 representing real kinetic  energy. \par 
  In Panels 3 and 4 of Figure 1 we show enlarged views of the small 
 sections in  the respective Panels 1 and 2  just to the right of the point $\kappa
 p=0$. Note the similarity between these energies even at this small 
 resolution and
 also note that despite this similarity the nonzero parts of the energy in Panel
 3 are about  $ 0.5\cdot 10^4 \ erg$ whereas those of Panel 4 are about $0.4 \ erg$. 
 The largest
 hooked negative graph corresponds to the smallest value of $c$ and as $c$
 increases  the
 other hooked positive and negative graphs are added. 
 The  larger
 $c$ becomes in Panels 3-4 the corresponding energies become smaller and  tend 
 to be densely 
 arrayed around zero.  This means that the  larger is the interval between
 the traps compared to their width the  kinetic energy of the diffusing
 particles tends to decrease to zero. \par
 \begin{figure}
\centerline{
\epsfxsize=6in
\epsffile{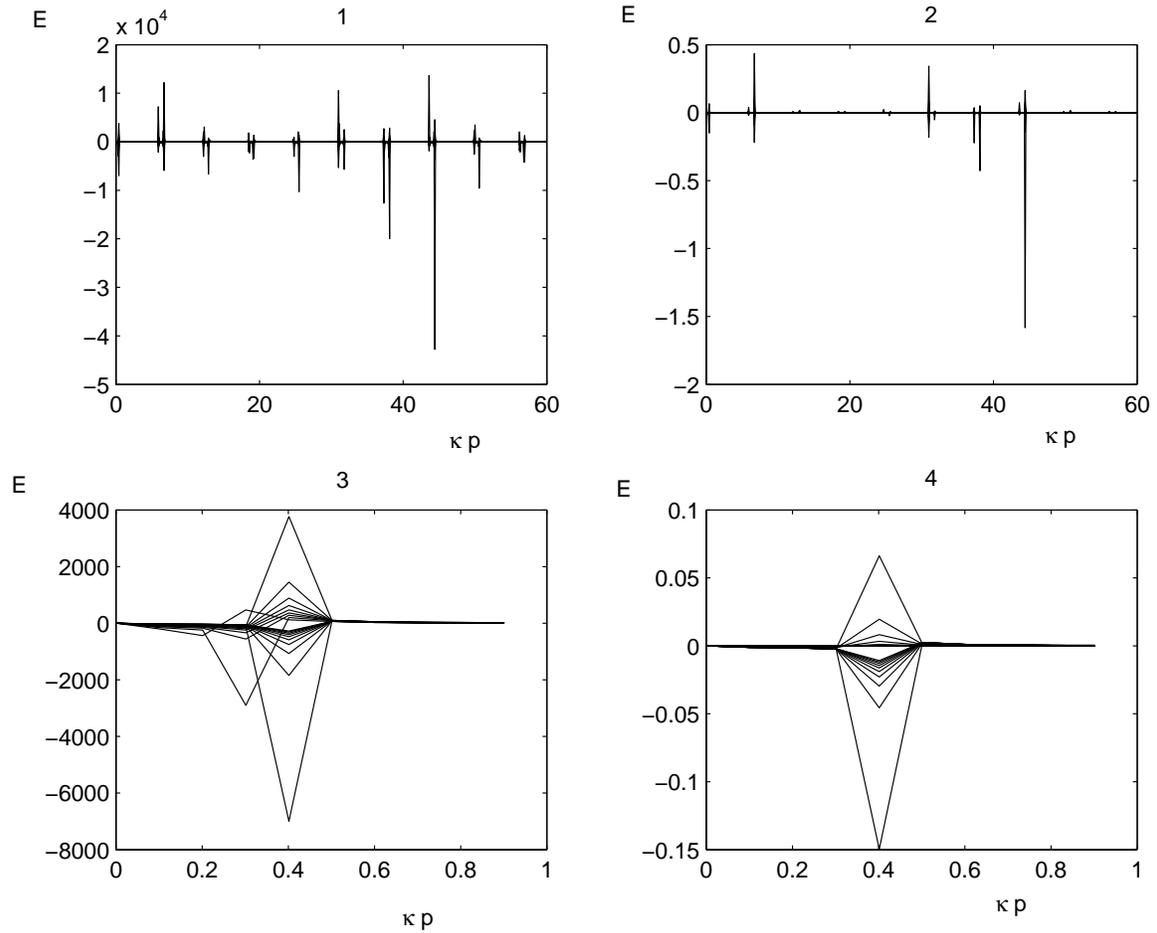}}
\caption{ Panels 1 and 2 respectively show the energies  $E^{-}(D_e,\frac{pc}{(1+c)},
 p,t)$ and $E^{+}(D_e,p,
 \frac{pc}{(1+c)},t)$  as functions of $\kappa p$ for the 20
 values of the ratio $c=0.1+n\cdot 0.4, \ \ \ n=0, 1, 2,\ldots 19$. Panels 3 and
 4 show a hight resolution of the respective  neighbourhouds in Panels 1 and 2 
 just to the right of the point $\kappa p=0$. Note that although Panels 1 and 2
 as well as Panels 3 and 4 are  similar in form  they greatly
 differ in the nonzero values of their energies. Both the trapping rate $k$ and
 the time $t$ have the
 values of  $k=t=2$  for all the graphs of  the four Panels. The units of the
 energies are in $ergs$. }
\end{figure} 
 The same similarity in graphical form and same large differences in  
 values may be
 shown for the same energies from Figure 1,    as functions of $\kappa p$,  but
 now for different
 values of the trapping rate $k$. This is seen in Panels 1 and 2 of Figure 2
 which respectively show the energies $E^{-}(D_e,\frac{pc}{(1+c)},
 p,t)$ and $E^{+}(D_e,p,
 \frac{pc}{(1+c)},t)$ as functions of $\kappa p$  using  
 the same 
 20 different values of  $k$  for each Panel and  the same
  $t$ and  $c$ for all the graphs shown. The 20 values of
 $k$ for each panel are $k=0.1 +n\cdot 0.4, \ \ \ n=1, 2, \ldots 19$ 
 and the values of  $t$ and
   $c$ for all the graphs  in the two Panels are 
 $t=2$ and $c=1$. As for Panels 1 and 2 of Figure 1 the differences between the
 nonzero parts of these energies amount to about $ 10^4$ although they look
 similar in external form.  Note that actually the energy in Panel 2 tends to
 zero. Panels 3 and 4 of Figure 2 respectively 
  show enlarged views
 of the respective neighbourhouds  from Panels 1 and 2 about 
 the point $\kappa p=25$.  One may note  the
 similarity between these energies even at this small resolution. Also, one may
 note that despite this apparent similarity the nonzero parts of the energy in
 Panel 3 is about $10^4$ whereas the corresponding ones in Panel 4 tend 
  to zero. \par 
 
\begin{figure}
\centerline{
\epsfxsize=6in
\epsffile{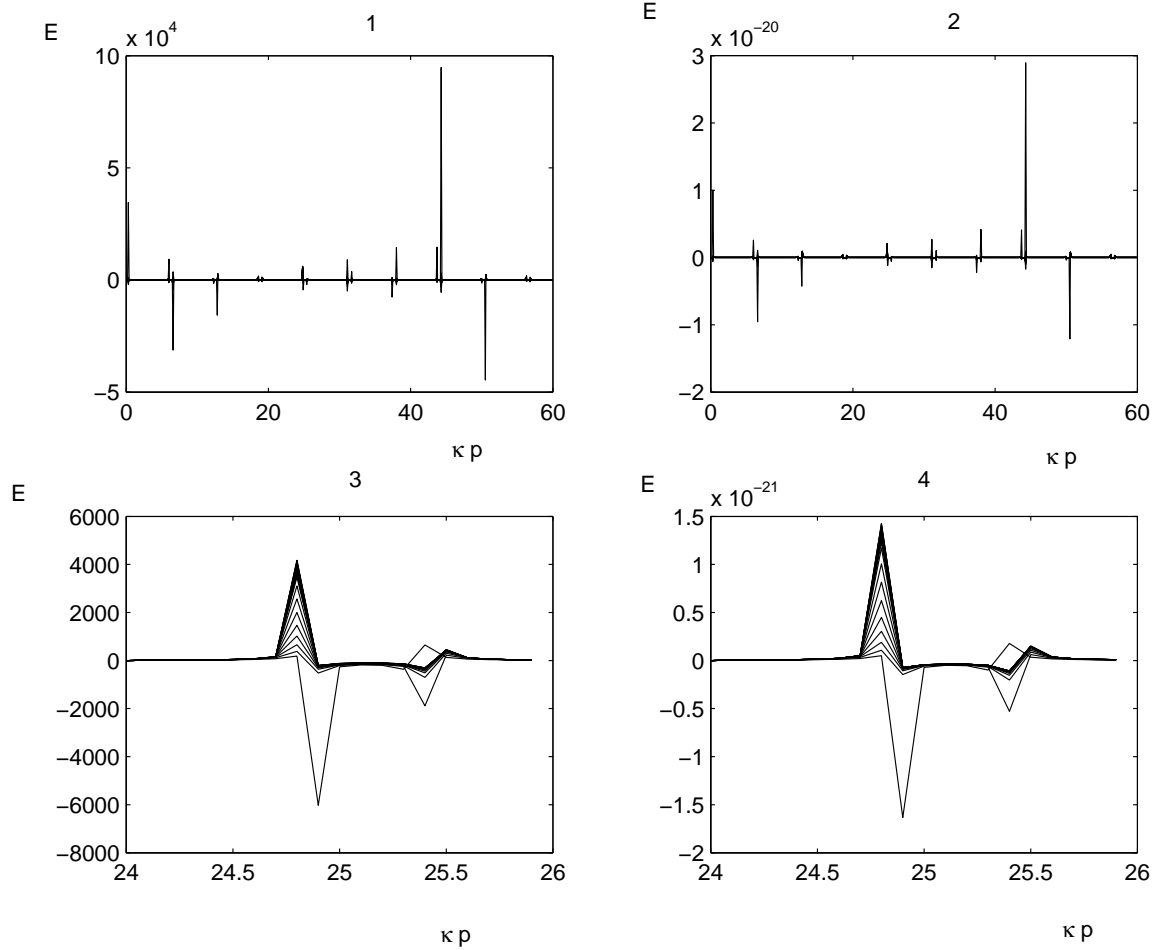}}
\caption{Panels 1 and 2 respectively show the energies  $E^{-}(D_e,\frac{pc}{(1+c)},
 p,t)$ and $E^{+}(D_e,p,
\frac{pc}{(1+c)},t)$  as functions of $\kappa p$ for the 20
 values of the trapping rate  $k=0.1+n\cdot 0.4, \ \ \ n=0, 1, 2,\ldots 19$.  
 Panel 3 and
 4 show a hight resolution of the respective  neighbourhouds in Panels 1 and 2 
 of  the point $\kappa p=25$. As is the case for the Panels of
 Figure 1 one may note  that although Panels 1 and 2
 as well as Panels 3 and 4 are very similar in form nevertheless they greatly
 differ in the nonzero values of their energies.  The ratio $c$ and
 the time $t$ have the respective 
 values of  $c=1$ and $t=2$   for all the graphs of  the four Panels. The
 energies are in units of $ergs$. }
\end{figure}
 For both Panels 3 and 4 the graphs with the large dense  hooked positive parts 
  correspond  to the smallest values of $k$ which means, as one may assume, that
  the smaller is the trapping rate of the traps the larger is the energy of the
  diffusing particles.   As $k$ increases  the corresponding graphs become 
   negative and they tend to zero for large enough $k$. 
    That is, the more $k$
 grows which means that the larger becomes the trapping rate of the trap 
 the more restrained and blocked become the diffusing particles in their passage
 through it. This is demonstrated through the vanishing of the positive allowed
 parts of the energies for large $k$  and their  tendency to the  zero  
  value. \par
 In the Panels of Figures 1-2 we compare for different values of $c$ and $k$ the
 two corresponding energies $E^{-}(D_e,\frac{pc}{(1+c)},
 p,t)$ and    $E^{+}(D_e,p,
\frac{pc}{(1+c)},t)$ as functions of $\kappa p$. We now 
 discuss   the second pair of corresponding energies 
 $E^{-}(D_e,p,
 p,t)$ and    $E^{+}(D_e,\frac{pc}{(1+c)},
 \frac{pc}{(1+c)},t)$.  Compared to the enrgies $E^{-}(D_e,\frac{pc}{(1+c)},
 p,t)$ and    $E^{+}(D_e,p,
 \frac{pc}{(1+c)},t)$  from Figures 1-2 the energies $E^{-}(D_e,p,
 p,t)$ and $E^{+}(D_e,\frac{pc}{(1+c)},
 \frac{pc}{(1+c)},t)$  are always positive and they are generally constant with 
 $\kappa p$.   Panels 1 and 2 respectively show 
 three-dimensional surfaces of 
the energies  $E^{-}(D_e,p,
 p,t)$ and $E^{+}(D_e,\frac{pc}{(1+c)},
 \frac{pc}{(1+c)},t)$  as functions of $\kappa p$ and $c$  and for the values of
  $k=t=2$. Note that the energy  $E^{+}(D_e,\frac{pc}{(1+c)},
 \frac{pc}{(1+c)},t)$ at Panel 2 does not depend at all on either  
  $\kappa p$ or
 $c$ and has the rather small constant value of $0.088 \ erg$.  
 The energy  $E^{-}(D_e,p,p,t)$ 
 at Panel 1  is constant for all $\kappa p$'s  and depends only slightly on 
 $c$ as
 seen from the small depression of the surface at small $c$ which causes it 
  to be
 slightly distorted from the planar form of Panel 2.  
 Panels 3 and
 4 respectively show  three-dimensional surfaces of 
the same energies  from Panels 1 and 2 but now   as functions 
of $\kappa p$ and $k$  and for the values 
  $c=1$ and $t=2$. Note that these energies, as in Panels 1 and 2, 
   do not depend  
  on $\kappa p$ and vary with $k$ to the maxima (for $c=1$)  of 
  $E_{max}^{-}(D_e,p,p,t) = 15 \ erg$ and $E_{max}^{+}(D_e,\frac{pc}{(1+c)},
 \frac{pc}{(1+c)},t)=0.23 \ erg$. Note also that as the trapping rate $k$  grows
  the energies 
  $E^{-}(D_e,p,p,t)$  and $E^{+}(D_e,\frac{pc}{(1+c)},
 \frac{pc}{(1+c)},t)$   decrease  in value and become zero at the  respective  
 values of  $k \approx 4.8$ and  $k \approx 6.3$. This result is  expected 
 since as
 the trapping rate grows the traps become more effective in blocking the
 diffusing particles.

   \par 
 \begin{figure}
\centerline{
\epsfxsize=5in
\epsffile{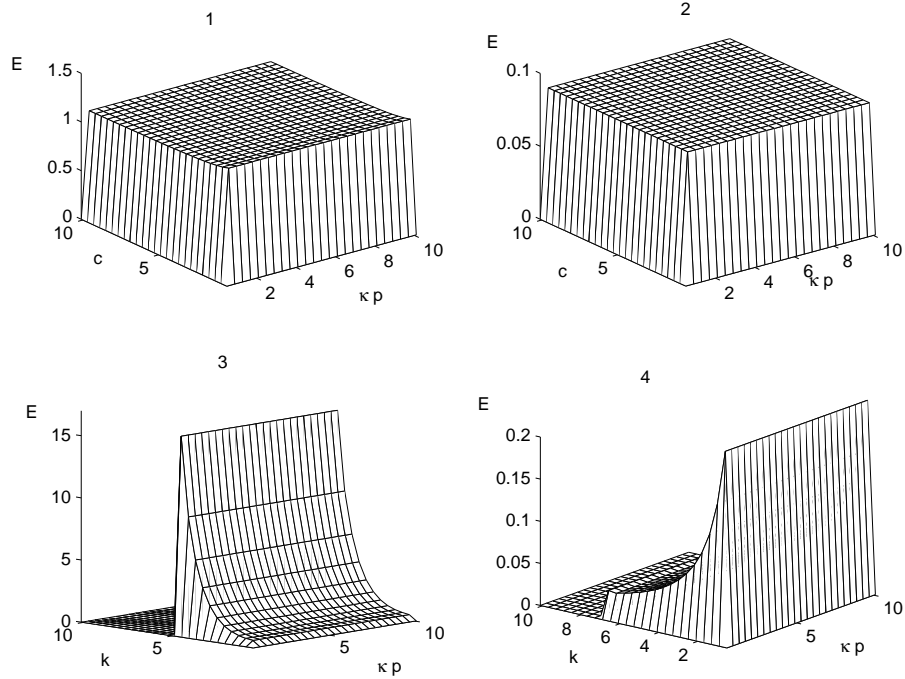}}
\caption{Panels 1 and 2 respectively show three-dimensional surfaces of 
the energies  $E^{-}(D_e,p,
 p,t)$ and $E^{+}(D_e,\frac{pc}{(1+c)},
 \frac{pc}{(1+c)},t)$  as functions of $\kappa p$ and $c$  and for 
  $k=t=2$. Note that the energy  $E^{+}(D_e,\frac{pc}{(1+c)},
 \frac{pc}{(1+c)},t)$ at Panel 2 does not depend on either   $\kappa p$ or
 $c$ and has the constant value of $0.088 \ erg$.  The energy  $E^{+}(D_e,p,p,t)$ 
 at Panel 1 depends only slightly  on $c$,   is constant 
 for all $\kappa p$'s  and
  have a maximum value (for $k=2$) of $E_{max}^{+}(D_e,p,p,t) = 1 \ erg$. 
 Panels 3 and
 4 respectively show  three-dimensional surfaces of 
the same energies  from Panels 1 and 2 but now   as functions 
of $\kappa p$ and $k$  and for 
  $c=1$ and $t=2$. Note that these energies do not depend at all 
  on $\kappa p$ and vary with $k$ to the maxima (for $c=1$)  of 
  $E_{max}^{-}(D_e,p,p,t) = 15 \ erg$ and $E_{max}^{+}(D_e,\frac{pc}{(1+c)},
 \frac{pc}{(1+c)},t)=0.23 \ erg$. Note also that the energies 
  $E^{-}(D_e,p,p,t)$  and $E^{+}(D_e,\frac{pc}{(1+c)},
 \frac{pc}{(1+c)},t)$    drop respectively 
 to zero at $k \approx 4.8$ and  $k \approx 6.3$.}
\end{figure}

 As seen from the Panels of Figures 1-3 
 the diffusing particles's energy considerably changes by merely 
 passing through the trap. Thus, referring to the pair  
 $E^{-}(D_e,\frac{pc}{(1+c)},
 p,t)$,   $E^{-}(D_e,p,
 p,t)$ one may realize  the
 large change in kinetic energy the diffusing particles goes through upon
 passing from the  left hand side to the right hand side of the trap. 
 For example,
 comparing Panel 1 of Figure 1, which shows the energy    
  $E^{-}(D_e,\frac{pc}{(1+c)},
 p,t)$ at the left hand side of the trap 
  to Panel 1 of Figure 3 which  shows the energy    
  $E^{-}(D_e,p,
 p,t)$ at the right hand side of this trap for
  the same values of $c$, $k$ and $t$ one may realize that 
 the particles's nonzero values of the energy changes upon diffusing through the trap 
 from $E \approx 10^4 \ erg$ to  $E \approx 1 \ erg$. These
 large differences may be realized again by comparing Panel 1 in Figure 2,  
 which shows the energy  
  $E^{-}(D_e,\frac{pc}{(1+c)},
 p,t)$ at the left hand side of the trap 
  to Panel 3 of Figure 3 which  shows the energy  
  $E^{-}(D_e,p,
p,t)$ at the right hand side of it  for
 the same  values of $k$, $c$ and $t$.  As seen,   the particles's nonzero
 values of the energy 
 decreases  upon passage of the trap from $E \approx 10^4$ to $E \approx 15$. 
 Thus, one may conclude that by diffusing through the traps the particles lose a
 huge amount of the energy they possess before the diffusion. \par  
 The time evolutions of the energies from Eqs (\ref{e31})-(\ref{e34}) as
 functions of $\kappa p$ reveal in a more pronounced way the mentioned large
 differences in the nonzero values of   the energies. This is demonstrated   in 
 the first two Panels of Figure 4
 which show the  energies $E^{-}(D_e,\frac{pc}{(1+c)},
 p,t)$  and   $E^{+}(D_e,p,
 \frac{pc}{(1+c)},t)$  for the 60 different values of  
 $t=1 +n\cdot 0.5, \ \ \ n=1, 2, \ldots 59$  in each panel and for 
   $c=2$ and $k=1$ 
   for all the graphs shown. 
   Note the giant differences of 
   about $10^{44}$ between  the
 nonzero values of the energies in Panels 1-2. Continuing to 
 increase $t$ causes  the energy in Panel 1 to grow (not shown) 
 even beyond $10^{80} \ erg$. Since these energies are not physically possible
 we conclude that there exist points along  the $\kappa p$ in 
 which the energies $E^{-}(D_e,\frac{pc}{(1+c)},p,t)$  are not allowed for large
 values of the time $t$. In Panel 3  we  show a 
 three-dimensional surface of
 the
 energies  $E^{-}(D_e,p,
 p,t)$  as function of $\kappa p$ and the time $t$. Note that it is constant
 with $\kappa p$ and decreases to zero, in contrast to the energy from 
 Panel 1, as $t$
 increases. Panel 4 shows the energy
 $E^{+}(D_e,\frac{pc}{(1+c)},\frac{pc}{(1+c)},t)$, as function of $\kappa p$, 
  for
 the 20 values of $t=1+n\cdot 0.5, \ n=0, 1, \ldots 19$ and for $c=2$ and
 $k=1$ for all the graphs. The dense line just above the abcissa axis denote the
 higher values of $t$ for which the constant values of the energy (as function
 of $\kappa p$) tend, like those of Panel 3,  to zero. \par
 From  the discussion thus far one may realize that generally the nonzero
 values of the energy  $E^{-}(D_e,\frac{pc}{(1+c)},p,t)$ are greater 
 by several order of
 magnitudes from the other three energies as shown by comparing the  Panels of 
 Figures 1, 2 and 4. These great differences are further pronounced for
 increasing values of the time as shown in the Panel 1 of Figure 3. 
 But that is no more so when the time decreases as turns out (not shown) 
 when the energies were calculated  at
 small times.  Thus, for example, decreasing the time from 
 $t \approx 30$ to $t \approx 0.2$ causes the nonzero values of the energy   
 $E^{-}(D_e,\frac{pc}{(1+c)},
 p,t)$  to decrease from  about $10^{44} \  erg$ (see Panels 1 of Figure 4) 
 to about $100 \ erg$.  Likewise, the energy    $E^{-}(D_e,p,
 p,t)$ decreases from about $15 \ erg$  (see Panel 3 of Figure 3) for $t=2$ 
 to $3 \ erg$ for $t=0.2$ (not shown). 

 \begin{figure}
\centerline{
\epsfxsize=6in
\epsffile{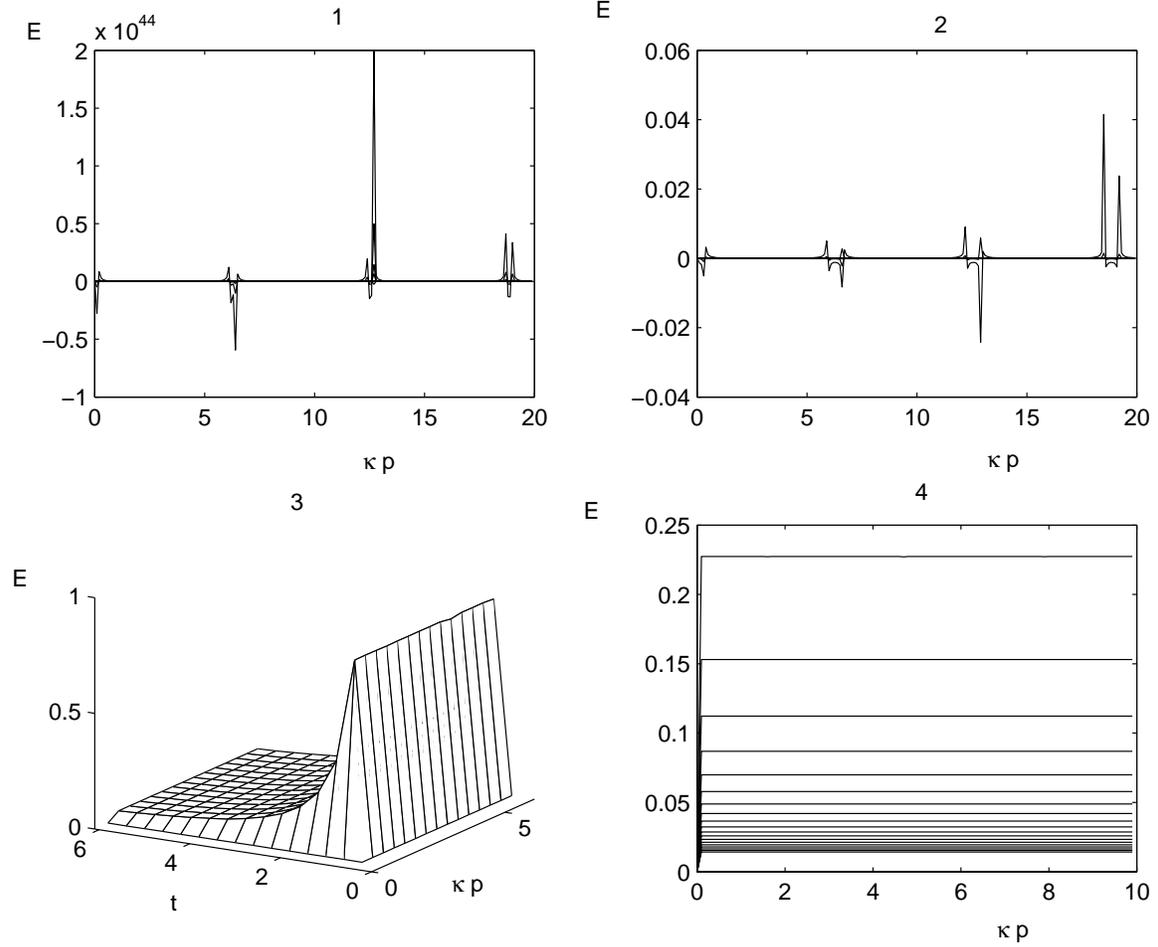}}
\caption{Panels 1 and 2 respectively show the energies  
$E^{-}(D_e,\frac{pc}{(1+c)},
p,t)$ and $E^{+}(D_e,p,
 \frac{pc}{(1+c)},t)$  as functions of $\kappa p$ for the 60
 values of the time  $t=1+n\cdot 0.5, \ \ \ n=0, 1, 2,\ldots 59$. One may
 realize that as the time grows the nonzero values of the energy  
 $E^{-}(D_e,\frac{pc}{(1+c)},
p,t)$    steeply increase.   
 The ratio $c$ and
 the trapping rate $k$ for all the graphs of Panels 1-2 are $c=2$ and $k=1$.  
 Compared to 
 the energy from Panel 1 which grows with $t$ the energies 
 $E^{-}(D_e,p,
p,t)$ and $E^{+}(D_e,\frac{pc}{(1+c)},
\frac{pc}{(1+c)},t)$  shown respectively in Panel 3 and
 4 decrease with time to zero.  Panel 4 is drawn for the 20 values of 
 $t=1+n\cdot 0.5, \ \ \ n=0, 1, 2,\ldots 19$ and for $c=2$ and $k=1$. 
 The upper lines in Panel 4 fit the small  values
 of $t$ and the lower lines fit the large values.  The energies  are given
 in units of $ergs$.}
\end{figure}

\protect \section{ Concluding Remarks}

We have discussed, using the transfer matrix method, 
   the energy of the particles which diffuse through the
unbounded one-dimensional multitrap system. The classical initial and boundary
value problem related to diffusion through an imperfect trap was adapted to
apply to an infinite array of similar traps as done in the sets 
(\ref{e1})-(\ref{e3}). Following the conventional transfer matrix procedure,
which is used for discussing the quantum Kronig-Penney multibarrier array, we
obtain a similar matrix equation (Eq (\ref{e8})) which relates the imperfect
traps across the whole array. Using, as for the analogous quantum multibarrier 
system, the periodicity of the array we obtain a quadratic characteristic
equation (Eq (\ref{e11})) for the two-dimensional matrix which relates the two
faces of the general $j-th$ trap. We solve this equation for the involved
eigenvalues of this matrix and impose upon them the finitness condition at the
limit at which the number $N$ of barriers becomes very large. As a result 
  two
inequalities (\ref{e16})-(\ref{e17}) are obtained which are the 
central expressions from
which we derive the appropriate kinetic energies of the diffusing particles.
Writing the matrix components $T_{11}(\frac{pc}{(1+c)},p)$ and  
$T_{22}(\frac{pc}{(1+c)},p)$  in these inequalities in terms of the appropriate
energies (see Eqs (\ref{e15})-(\ref{e23}) and discussion there) and using the
properties of the transfer matrix method (see Eqs (\ref{e18}), (\ref{e21}) and
(\ref{e24})) we obtain two simultaneous equations involving two energies. We
have found that        
each of these
two energies is composed of  two parts; one is  related to the left hand 
 face of the
trap and the second to its right hand face. It is also  found that the  two 
parts of each of the two  energies differ
greatly from each other not only in value but also in 
 the way they are  expressed  as functions of the
related variables $c$, $k$, $t$ and $\kappa p$. That is, as  seen from 
 Eqs  (\ref{e31})-(\ref{e34})  one part of each  of these two energies
is expressed in  ideal trap terms only and the second part in imperfect trap 
terms only.  
These differences entail the results that 
 by merely diffusing through the trap the particles's energy totally changes as
 realized from the appended figures. Moreover, there exist great variations 
  not
 only between  the two parts  of the same energy  but
 also between different sections (along the $\kappa p$ axis) 
  of the same part itself.  For example, in the respective Panels 1 of Figures 
  1, 2 and  4 we have
 found that the energy $E^{-}(D_e,{\grave
x_j}^{left},
 p,t)$ at the left hand face of the trap assumes  values which greatly varies
 even in  very short ranges of $\kappa p$.  \par 
 As  discussed in Section IV  the points along 
 the $\kappa p$ axis 
 at which the energies may assume unexpected,  or even disallowed  values,   
  are related to 
  the following two kinds of points; (1) 
   $\kappa p=\frac{\pi}{2}+n\pi, \ \ n=0, 1, 2, \ldots$ \ \ \ 
 (2) $\kappa p=arc(\cos(\frac{A({\grave x_j}^{right},D_e)}{A({\grave
x_j}^{left},D_e)}))$. Another important variable which entails a large changes
in the values of the energies is the time $t$ as realized from the  
Panels
of the appended Figures (see, especially, Panels 1-4 of Figure 4). The
analytical expressions obtained are corroborated by  the different attached 
figures. \par
As noted, the analogous discussion of the quantum Kronig-Penney multibarrier   
entails the finding of points along the corresponding $\kappa p$ axis at which
the energy is disallowed. Here, for the classical imperfect multitrap we have
found corresponding disallowed energies which take the form of either negative
values for the kinetic energy or of a discontinuous change of this energy from zero
to  enormous positive values as in Panel 1 of Figure 4. The quantum band-gap
structure found in the Kronig-Penney multibarrier array have been turned out to have 
great applications in wide areas of solid state physics such as semiconductor
devices and computer chips. The striking similarity in the forms  of the
Schroedinger and diffusion equations as well as the common possibility to
investigate and discuss them by the transfer matrix method may entail in 
the future similar
successful development for the classical diffusive systems.  

\appendix

\underline{\large \bf APPENDIX A }

\vspace{0.2 cm}

 \section{ The  matrix elements  from Eq (\ref{e7})}

  The  matrix elements 
$T_{11}({\grave x_j}^{left},{\grave x_j}^{right})$, 
$T_{12}({\grave x_j}^{left},{\grave x_j}^{right})$, 
$T_{21}({\grave x_j}^{left},{\grave x_j}^{right})$, and 
$T_{22}({\grave x_j}^{left},{\grave x_j}^{right})$ of the 
two-dimensional matrix $T^{(j)}$ from Eq (\ref{e7}) are fully discussed and
derived in \cite{Bar1,Bar2}  and are given by the following expressions
$$    T_{11}({\grave x_j}^{left},{\grave x_j}^{right})=
\frac{\alpha(D_e,{\grave x_j}^{left},{\grave x_i},t)
\alpha(D_i,{\grave x_j}^{right},{\grave x_i},t)}{\alpha(D_i,{\grave
x_j}^{left},{\grave x_i},t)
\alpha(D_e,{\grave x_j}^{right},{\grave x_i},t)},  
 \ \ \ 1 \le j \le N      \eqno(A_1) $$    
$$   T_{12}({\grave x_j}^{left},{\grave
x_j}^{right})=0,  
\ \ \ 1 \le j \le N       \eqno(A_2) $$ 
$$
   T_{21}({\grave x_j}^{left},{\grave x_j}^{right})=
\rho_0(\frac{\eta(D_i,{\grave x_j}^{right},t)}
{\eta(D_e,{\grave x_j}^{right},t)}(\frac{\xi(D_e,{\grave x_j}^{left},{\grave x_i},t)}
{\eta(D_i,{\grave x_j}^{left},t)} - $$ 
$$- \frac{\alpha(D_e,{\grave x_j}^{left},{\grave
x_i},t)\xi(D_i,{\grave x_j}^{left},{\grave x_i},t)}
{\alpha(D_i,{\grave x_j}^{left},{\grave x_i},t)\eta(D_i,{\grave x_j}^{left},t)})) + 
$$   $$
+\frac{\alpha(D_e,{\grave x_j}^{left},{\grave x_i},t)}
{\alpha(D_i,{\grave x_j}^{left},{\grave x_i},t)}(\frac{\xi(D_i,{\grave
x_j}^{right},{\grave x_i},t)}
{\eta(D_e,{\grave x_j}^{right},t)} - \eqno(A_3) $$ 
$$ - \frac{\alpha(D_i,{\grave x_j}^{right},{\grave
x_i},t)\xi(D_e,{\grave x_j}^{right},{\grave x_i},t)}
{\alpha(D_e,{\grave x_j}^{right},{\grave x_i},t)\eta(D_e,{\grave x_j}^{right},t)}), 
\ \ \ 1 \le j \le N $$ 
$$  T_{22}({\grave x_j}^{left},{\grave
x_j}^{right})=\frac{\eta(D_e,{\grave x_j}^{left},t)
\eta(D_i,{\grave x_j}^{right},t)}{\eta(D_i,{\grave x_j}^{left},t)
\eta(D_e,{\grave x_j}^{right},t)}, \ \ \ 1 \le j \le N  \eqno(A_4) $$

The parameters $\alpha$, $\xi$, and $\eta$ are given by (we write these expression for 
$D_e$ and $x={\grave x_j}^{left}$) 
$$  \alpha(D_e,{\grave x_j}^{left},{\grave x_i},t)=
erf(\frac{({\grave x_j}^{left}-{\grave x_i})}{2\sqrt{D_et}})+
\exp(k^2D_et+k({\grave x_j}^{left}-{\grave x_i}))\cdot $$
  $$ \cdot erfc(k\sqrt{D_et}+ 
\frac{({\grave x_j}^{left}-{\grave x_i})}{2\sqrt{D_et}}) \eqno(A_5) $$
 
$$  \xi(D_e,{\grave x_j}^{left},{\grave x_i},t)=
k\exp(k^2D_et+k({\grave x_j}^{left}-{\grave x_i}))
erfc(k\sqrt{D_et}+\frac{({\grave x_j}^{left}-{\grave x_i})}{2\sqrt{D_et}}) 
\eqno(A_6)   $$ 
$$ \eta(D_e,{\grave x_i},t) =-\frac{\pi}{{\grave x_i}}
e^{-(\frac{\pi}{{\grave x_i}})^2D_et} \eqno(A_7)   $$
Note that in \cite{Bar1,Bar2} the variables ${\grave x_i}$ are not subtracted
from the variables ${\grave x_j}$ in the functions $\alpha$ and $\xi$. This is
because the presence or absence of this subtraction do not affect at all the
values of the matrix elements $T_{11}$ and $T_{22}$ as may be realized from
their definitions in Eqs $(A_1)$ and $(A_4)$ in this Appendix.  
 Also,  in \cite{Bar1,Bar2} we discuss 
 the whole array of the bounded dense  multitrap in which case the variables 
 ${\grave x_i}$  and ${\grave x_j}$ do not, necessarily, refer to the same trap 
 and so this subraction is ignored there.  Here, on the other hand, 
 the variables 
 ${\grave x_i}$  and ${\grave x_j}$ refer to the same trap which 
 represents   the unbounded multitrap system and so 
the expression  
$({\grave x_j}-{\grave x_i})$   should not be approximated to ${\grave x_j}$. 
  
\newpage

\underline{\large \bf APPENDIX B} 

\vspace{0.2 cm}

\protect \section{The solutions of the simultaneous Eqs (\ref{e25})-(\ref{e26}) }

We solve in this Appendix the two Eqs (\ref{e25})-(\ref{e26}) for the energies
$E(D_e,\frac{pc}{(1+c)},{\grave x_i},t)$ and  
 $E(D_e,p,{\grave x_i},t)$. We begin by solving  Eq  
 (\ref{e25})
for $E(D_e,\frac{pc}{(1+c)},{\grave x_i},t)$  in terms of 
$E(D_e,p,{\grave x_i},t)$ and then we substitute this
solution in Eq (\ref{e26}) and  solve it for 
$E(D_e,p,{\grave x_i},t)$. In order not to be involved with 
cumbersome expressions we define the following quantities  
 $$
 X_1=\frac{A(\frac{pc}{(1+c)},D_e)}{A(p,D_e)}\cdot
 \cos(\kappa p) $$
 $$X_2=D_eB(\frac{pc}{(1+c)},D_i)\rho_2(D_i,\frac{pc}{(1+c)},{\grave
 x_i},t)  $$
 $$X_3=D_eB(\frac{pc}{(1+c)},D_e)\rho_2(D_e,\frac{pc}{(1+c)},{\grave
 x_i},t)  \eqno(B_1)  $$
 $$X_4=D_eB(p,D_e)\rho_2(D_e,p,{\grave
 x_i},t)  $$
 $$X_5=D_eB(p,D_i)\rho_2(D_i,p,{\grave
 x_i},t)  $$ 
 Substituting the last quantities in Eq (\ref{e25}) we obtain 
 $$  
 t^2(1-X_1)E(D_e,\frac{pc}{(1+c)},{\grave x_i},t)
 E(D_e,p,{\grave x_i},t)-
 t\biggl(E(D_e,\frac{pc}{(1+c)},{\grave x_i},t)(X_5-X_4X_1)+ $$
 $$ +
 E(D_e,p,{\grave x_i},t)(X_3-X_2X_1)\biggr)+
 X_3X_5-X_2X_4X_1=0  \eqno(B_2) $$ 
 Solving the last equation for $E(D_e,\frac{pc}{(1+c)},{\grave x_i},t)$ 
 we obtain
 $$ E(D_e,\frac{pc}{(1+c)},{\grave
 x_i},t)=\frac{X_1X_2(X_4-tE(D_e,p,{\grave
 x_i},t))-X_3(X_5-tE(D_e,p,{\grave x_i},t))}
 {tX_1(X_4-tE(D_e,p,{\grave
 x_i},t))-t(X_5-tE(D_e,p,{\grave x_i},t))} \eqno(B_3)  $$
 
 We may now substitute the last expression for 
$E(D_e,\frac{pc}{(1+c)},{\grave x_i},t)$ in Eq (\ref{e26}) and solve it 
for $E(D_e,p,{\grave x_i},t)$. But before proceeding we
define the following quantities 
$$ Y_1=t^2(\frac{B(p,D_e)}
{B(\frac{pc}{(1+c)},D_e)}- \cos(\kappa p)) $$ 
 $$Y_2=tD_e\left(\frac{B(p,D_e)}
{B(\frac{pc}{(1+c)},D_e)}A(p,D_i)
\alpha(D_i,p,{\grave x_i},t) - A(p,D_e)
\alpha(D_e,p,{\grave x_i},t)\cos(\kappa p)\right) $$
$$ Y_3=tD_e\biggl(\frac{B(p,D_e)}
{B(\frac{pc}{(1+c)},D_e)}A(\frac{pc}{(1+c)},D_e)
\alpha(D_e,\frac{pc}{(1+c)},{\grave x_i},t) - A(\frac{pc}{(1+c)},D_i) \cdot
$$   $$ \cdot 
\alpha(D_i,\frac{pc}{(1+c)},{\grave x_i},t)\cos(\kappa p)\biggr) \eqno(B_4) $$ 
 $$ Y_4=D_e^2\frac{B(p,D_e)}
{B(\frac{pc}{(1+c)},D_e)}A(p,D_i)
\alpha(D_i,p,{\grave x_i},t) A(\frac{pc}{(1+c)},D_e)
\alpha(D_e,\frac{pc}{(1+c)},{\grave x_i},t) $$
$$ Y_5=D_e^2A(p,D_e)
\alpha(D_e,p,{\grave x_i},t) A(\frac{pc}{(1+c)},D_i)
\alpha(D_i,\frac{pc}{(1+c)},{\grave x_i},t)\cos(\kappa p) $$  

Substituting from Eqs $(B_3)$-$(B_4)$ in Eq (\ref{e26}) and rearranging 
we obtain the
following  quadratic equation for $E(D_e,p,{\grave
 x_i},t)$ 
 $$
 E^2(D_e,p,{\grave
 x_i},t)\biggl((tX_3-tX_1X_2)Y_1-t^2(1-X_1)Y_3\biggr)+
 E(D_e,p,{\grave
x_i},t)\cdot $$   $$ \cdot 
\biggl\{(t^2(1-X_1)(Y_4-Y_5)-(tX_4X_1-tX_5)Y_3-t(X_3-X_1X_2)Y_2+
(X_4X_1X_2- $$     $$ -X_3X_5)Y_1\biggr\}-
(X_4X_1X_2-X_3X_5)Y_2+t(X_4X_1-X_5)(Y_4-Y_5)=0
\eqno(B_5)  $$ 
  The two solutions  of the  last quadratic equation  are 
$$ E^{+}(D_e,p,{\grave
x_i},t)=\frac{t(X_1X_4-X_5)Y_3-(X_1X_2X_4-X_3X_5)Y_1}
{(tX_3-tX_1X_2)Y_1-t^2(1-X_1)Y_3} \eqno(B_6) $$ 
$$ E^{-}(D_e,p,{\grave
x_i},t)=\frac{(X_3-X_1X_2)Y_2-t(1-X_1)(Y_4-Y_5)}
{(X_3-X_1X_2)Y_1-t(1-X_1)Y_3}=\frac{Y_2}{Y_1},
\eqno(B_7) $$ 
where the last result for $ E^{-}(D_e,p,{\grave
x_i},t)$ is obtained by using Eqs $(B_4)$. 
The  two expressions from $(B_6)$-$(B_7)$  are the energies at the 
right hand side of the trap. 
The corresponding energies $E^{\pm}(D_e,\frac{pc}{(1+c)},{\grave
x_i},t)$ at the left hand side of it may be obtained from Eq $(B_3)$ 
by substituting in it for $E^{\pm}(D_e,p,{\grave
x_i},t)$ from Eqs $(B_6)$-$(eB_7)$. Thus, using Eq $(B_6)$ one may
find $E^{+}(D_e,\frac{pc}{(1+c)},{\grave
x_i},t)$ as 
$$   E^{+}(D_e,\frac{pc}{(1+c)},{\grave
x_i},t)=\frac{Y_3}{Y_1} \eqno(B_8)  $$
where use is made of the first and third  of Eqs $(B_4)$. 
The second energy  $E^{-}(D_e,\frac{pc}{(1+c)},{\grave
x_i},t)$ is obtained by substituting from Eq $(B_7)$ in Eq $(B_3)$. 
$$  E^{-}(D_e,\frac{pc}{(1+c)},{\grave
x_i},t)= \eqno(B_9) $$
$$=\frac{ \scriptstyle (X_3-X_1X_2)((tX_3-tX_1X_2)Y_2+
(X_1X_2X_4-X_3X_5)Y_1)-(1-X_1)(t^2(X_3-X_1X_2)(Y_4-Y_5)+t(X_1X_2X_4-X_3X_5)Y_3)}
{ \scriptstyle
(X_3-X_1X_2)((t^2(1-X_1)Y_2+t(X_1X_4-X_5)Y_1)-(1-X_1)(t^3(1-X_1)(Y_4-Y_5)+
t^2(X_1X_4-X_5)Y_3)} 
$$

\bigskip \bibliographystyle{plain}

\end{document}